\documentclass{ptptex}
\usepackage{amssymb,amsfonts,amsmath}
\usepackage{latexsym}
\usepackage[dvips]{graphicx}

\newcommand{\spar}{\partial \hspace{-2.9mm}{\not}\hspace{2.9mm}}
\newcommand{\slaA}{A \hspace{-2.9mm}{\not}\hspace{2.9mm}}
\def\nn{\nonumber}

\title{Four-Fermion Interaction Model in a Constant Magnetic Field
\\ at Finite Temperature and Chemical Potential
}
\author{Tomohiro \textsc{Inagaki}${}^{1,}$
\footnote{inagaki@hiroshima-u.ac.jp},
Daiji \textsc{Kimura}${}^{2,}$   
\footnote{kimura@theo.phys.sci.hiroshima-u.ac.jp}
\ and Tsukasa \textsc{Murata}${}^{2,}$ 
\footnote{murata@theo.phys.sci.hiroshima-u.ac.jp}}
\inst{
${}^1$Information Media Center, Hiroshima University, 
Higashi-Hiroshima \\
739-8521, Japan \\
${}^2$Department of Physics, Hiroshima University, 
Higashi-Hiroshima\\
739-8526, Japan}

\abst{
We investigate the influence of an external magnetic field on chiral
symmetry breaking in a four-fermion interaction model at finite
temperature and chemical potential. Using the Fock-Schwinger proper time
method, we calculate the effective potential for the four-fermion
interaction model to leading order in the $1/N_c$ expansion. The phase
structure of the chiral symmetry breaking is determined in the
$T$-$\mu$, $H$-$T$ and $\mu$-$H$ planes. The external magnetic field
modifies the phase structure. It is found that a new phase appears for a
large chemical potential.
}
\begin{document}
\maketitle

\section{Introduction}

The strong interaction between quarks and gluons is described by QCD. In
recent years, much interest has been paid to the phase structure of
QCD. The Lagrangian of QCD has a $SU(N)_L\otimes SU(N)_R$ symmetry for
massless quarks. The chiral $SU(N)_L \otimes SU(N)_R$ symmetry is 
broken down to the flavor $SU(N)$ symmetry by the QCD dynamics. It is
expected that the chiral symmetry is restored at high temperature and/or
high density. A phase transition takes place from hadronic matter to a
quark-gluon plasma or a color super-conducting phase under hot and dense
conditions. This affects the structure of spacetime and the evolution of
stars. 

A state of high temperature and density was realized in the early
states of our universe. There is a possibility that the early universe 
contained a large primordial magnetic field. Another dense state at low 
temperature is found in the cores of neutron stars. Some neutron stars, 
called pulsars, possess a strong magnetic field and emit high energy
radiation. The existence of a so-called `magnetar', which is a star with
a much stronger magnetic field, has been reported. To describe such
state, it is very important to investigate thermal and magnetic effects
on the QCD ground state. To study dense states, we introduce a chemical 
potential. The phase structure of QCD is studied in an external magnetic 
field $H$ at finite temperature $T$ and chemical potential $\mu$.

Because a long-range correlation plans an essential role in the phase 
transition, the QCD ground state is determined non-perturbatively.
There are many methods to deal with non-perturbative aspects of QCD,
e.g. lattice QCD, the Schwinger-Dyson equation, the strong coupling
expansion, and so on. In the present paper, we consider a four-fermion
interaction model as a low-energy effective theory of QCD.\cite{Namb} It
is known that this model has properties similar to those of
non-perturbative QCD at the low energy regime. In the model, chiral
symmetry is dynamically broken down through fermion and anti-fermion
condensation for a sufficiently strong four-fermion interaction. 

Employing the Matsubara formalism of finite temperature field theory,
the thermal effect on chiral symmetry breaking was investigated in the
Nambu-Jona-Lasinio (NJL) model.\cite{Hatsu} For a high temperature
and/or a large chemical potential, the expectation value of a composite
state consisting of a fermion and an anti-fermion,
$\langle\bar{\psi}\psi\rangle$, disappears, and broken chiral symmetry
is restored. A tricritical point between the first and second order
phase transitions was observed in the critical curve in the $T$-$\mu$
plane for $H=0$ \cite{Klev,Chod,Ina2}.  

A method to treat a strong external magnetic field was developed by
Schwinger in 1951\cite{Schw}. This method is called the Fock-Schwinger
proper time method. In Refs. \citen{Suga} and \citen{Mira:1994PRL}, the
proper time method is applied to a four-fermion interaction model to 
study the influence of a magnetic field. It is shown that an external
magnetic field has the effect of effectively reducing the number of
spacetime dimensions and enhancing the chiral symmetry. Thus, the
expectation value $\langle\bar{\psi}\psi\rangle$ slightly increases in a
magnetic field at $T=\mu=0$. This method has been extended to a variety
of situations with external fields.\cite{Chiu,Ina,Kane,Eber1} 

Here we apply the Fock-Schwinger proper time method to the NJL model
at finite temperature, chemical potential and external magnetic
field. Using an explicit expression of the fermion Green function in an
external magnetic field at finite $T$ and $\mu$ \cite{IKM}, we calculate
the effective potential to leading order in the $1/N_c$-expansion. We
use the observed quantities $f_\pi$ and $m_\pi$ to fix the parameters in
the model. The ground state is determined by observing the minimum of
the effective potential. Evaluating it numerically, we determine the
phase structure of the theory in the $T$-$\mu$, $H$-$T$ and $\mu$-$H$
planes and investigate the combined effects of the magnetic field,
temperature and chemical potential. Finally, we give some concluding
remarks.  

\section{Effective potential with $T$, $\mu$ and $H$}

In this section, we briefly review the fermion Green function at 
finite $T$ and $\mu$ with an external magnetic field and calculate the
effective potential for a four-fermion interaction model to leading
order in the $1/N_c$ expansion. There are some
works\cite{Klev,Elmfors:1996fx,Schm,Ditt} in which the original
Fock-Schwinger proper time method\cite{Schw,Itzy} is applied to finite
temperature and/or chemical potential situations. It has been pointed
out that a naive Wick rotation is not always valid for a proper time
integral. In order to investigate physical situations, we carefully
choose the proper time contour, as demonstrated in Ref. \citen{IKM}. We
apply this result to the calculation of the effective potential in the
four-fermion interaction model. 

Here we consider the NJL model, which is well known as one of the low
energy effective theories, in the chiral limit of QCD. The Lagrangian of
the NJL model is given by  
\begin{equation}
 {\cal L} = \sum_{i=u,d}\bar{\psi}_i (i\spar 
       -Q_i A\hspace{-2.9mm}{\not}\hspace{2.9mm}) \psi_i
       + \frac{G}{2N_c}
        \left\{ \left( \sum_{i=u,d} \bar\psi_i \psi_i \right)^2 
         + \left( \sum_{i,j=u,d}\bar\psi_i i\gamma_5 
                    \vec{\tau}_{ij}\psi_i \right)^2 \right\} \ ,
\label{NJL-lag}
\end{equation}
where $N_c$ is the number of colors, the indices $i$ and $j$ denote the
fermion flavors, $Q_i$ represents the electric charge of the quark
fields ($Q_u=2e/3$, $Q_d=-e/3$), and $G$ is the effective coupling
constant. We consider two flavors of quark fields. The quark fields
$\psi_i$ belong to the fundamental representation of the color $SU(N_c)$
group and a flavor isodoublet. $\vec{\tau}$ represents the isospin Pauli
matrices. This Lagrangian is invariant under the global chiral
transformation $\psi_i\to \exp\left\{i\theta\frac{\tau^3_{ij}}2\gamma_5
\right\}\psi_j$.\footnote{Because the electric charges for up and down
quarks are different, the flavor $SU(N)$ symmetry is broken down
explicitly. Only the symmetry under this transformation remains.}
Below, we do not include the flavor indices explicitly. In practical
calculations, it is more convenient to introduce the auxiliary field
$\sigma$ and $\vec{\pi}$, and hence we consider the Lagrangian 
\begin{equation}
 {\cal L} 
  = \bar{\psi}\left( i\spar - QA\hspace{-2.9mm}{\not}\hspace{2.9mm}
        - \sigma - i\gamma_5 \vec{\tau}\cdot\vec{\pi} \right) \psi
       +\frac{N_c}{2G}
          \left( \sigma^2 + \vec{\pi}^2 \right) .
\label{aux-lag}
\end{equation}
From the equations of motion for $\sigma$ and $\vec{\pi}$, we obtain the
correspondences $\sigma \sim -(G/N_c) \bar\psi \psi$ and $\vec{\pi} \sim
-(G/N_c) \bar\psi i\gamma_5 \vec{\tau} \psi$. 

To find the ground state of the NJL model at finite $T$, $\mu$ and $H$,
we evaluate the effective potential. Following the standard procedure of
the imaginary time formalism\cite{LeB}, we introduce the temperature and
the chemical potential. Here, we assume that the ground state does not
break the local $U(1)_{EM}$ symmetry, i.e. $\left< \pi^+ \right>=\left<
\pi^-\right>=0$. Under this assumption, we can choose the isospin
singlet ground state from the degenerate state using the chiral
transformation $\psi\to \exp\left\{i\theta\frac{\tau^3}2\gamma_5
\right\}\psi$. To leading order in the $1/N_c$ expansion, the effective
potential $V_{\rm eff}(\sigma,\vec\pi=0)$ is given by 
\begin{equation}
V_{\rm eff}(\sigma,\vec\pi=0)
  = \frac{ 1 }{4 G} \sigma^2 
     -\frac{1}{2\beta V}{\rm Tr}\ln{\left( 
   \frac{i\spar + Q\slaA -i\mu\gamma_4 -\sigma}
            {i\spar + Q\slaA -i\mu\gamma_4} \right)} ,
\label{v-fer0}
\end{equation}
where we assume $\mu=\mu_u=\mu_d$ and ${\rm Tr}$ denotes the trace over
spacetime coordinates, spinor and flavor indices. The factor $\beta V$
in the denominator comes from  the 4-dimensional volume,
\begin{equation}
         \int^\beta_0 \! dx_4 \int_{-\infty}^\infty \! d^3{\vec{x}} 
         = \beta V ,
\end{equation}
where $\beta=1/(k_B T)$, with $k_B$ the Boltzmann constant. Below, we
set $k_B =1$. The effective potential (\ref{v-fer0}) is normalized to
satisfy $V_{\rm eff}(\sigma=0)=0$.  

The second term on the right-hand side of Eq. (\ref{v-fer0}) is
rewritten using the relation 
\begin{equation}
{\rm Tr}\ln{\left(
   \frac{i\spar + Q\slaA -i\mu\gamma_4 -\sigma}
         {i\spar + Q\slaA -i\mu\gamma_4}  \right)}
    = {\rm Tr}\int_0^\sigma S(x,x;m) dm , 
\label{v-fer}
\end{equation}
where $S(x,y;m)$ is the fermion Green function, which satisfies the
Dirac equation  
\begin{eqnarray}
 \left( i\spar + Q\slaA -i\mu\gamma_4 -m \right) S(x,y;m)
        = \delta^4_{\rm E}(x-y) \ ,
\label{Green-fer}
\end{eqnarray}
where $\delta^4_{\rm E}(x-y) = \delta(x_4-y_4)\delta^3(\vec{x}-\vec{y})$. 

To solve the Dirac equation (\ref{Green-fer}), we introduce a bispinor
function $G(x,y;m)$, defined by
\begin{equation}
 S(x,y;m) =
 \left( i\spar + Q\slaA -i\mu\gamma_4 +m \right) G(x,y;m) \ .
\label{Green-G}
\end{equation}
Substituting Eq. (\ref{Green-G}) into Eq. (\ref{Green-fer}), we obtain
\begin{equation}
 \left\{ \sum_{j=1}^3 (\partial_j -iQA_j)^2 
       +\frac12 Q F_{jk}\sigma^{jk}
        -(i\partial_4-i\mu)^2 -m^2 \right\}G(x,y;m) 
	=\delta^4_{\rm E}(x-y) ,
\label{eq-Gre}
\end{equation}
where $F_{ij}$ is the three-dimensional field strength and 
$\sigma^{jk}=(i/2)[\gamma^j,\gamma^k]$.  

Performing the Fourier series expansion,
\begin{equation}
 G(x,y;m) 
   = \frac1\beta \sum^{\infty}_{n=-\infty}
       e^{-i\omega_n (x_4-y_4)} \widetilde{G}_n(\vec{x}, \vec{y};m) \quad,
 \quad \omega_n = \frac{(2n+1)\pi}\beta ,
\label{Til-G}
\end{equation}
we rewrite Eq. (\ref{eq-Gre}) into the form
\begin{eqnarray}
&& \biggl\{ \sum_{j=1}^3\left(\partial_j -iQA_j \right)^2
          +\frac{QH}2 (\gamma_1\gamma_2-\gamma_2\gamma_1)
 \biggr. \nonumber \\ && \hspace{4.5cm} \biggl. 
    -(\omega_n -i\mu)^2
  - m^2 \biggr\} \widetilde{G}_n(\vec{x},\vec{y};m) 
      = \delta^3(\vec{x} - \vec{y}) ,
\label{Fou-G}
\end{eqnarray}
where we consider a constant magnetic field along the $x_3$-direction,
i.e. $F_{12}=-F_{21}=H$, for simplicity.\footnote{If we consider a
constant electric field, we obtain an equation similar to
Eq. (\ref{Fou-G}), except for a $\vec{x}$-dependent potential. Due to
the presence of this $\vec{x}$-dependent term, we cannot analytically
solve this equation in an external electric field at finite $T$ and
$\mu$.} 

As is well known, the solution of Eq. (\ref{Fou-G}) can be obtained
using the Fock-Schwinger proper time method\cite{Schw,Itzy}. In
Ref. \citen{Elmfors:1996fx}, it is pointed out that the original
Fock-Schwinger method must be modified in the case that $T\not=0$ and
$\mu\not=0$. The function $\widetilde{G}_n(\vec{x},\vec{y};m)$ is
determined by  
\begin{eqnarray}
 \widetilde{G}_n(\vec{x},\vec{y};m) = \left\{
\begin{array}{ll}
 e^{-5\pi i/4} \int^{-0}_{-\infty}   U_n(\vec{x},\vec{y};\tau)~d\tau 
      & \mbox{for ~$n \geq0$}\\
 e^{5\pi i/4}  \int_{+0}^{\infty} U_n(\vec{x},\vec{y};\tau)~d\tau  
      & \mbox{for ~$n<0$}
\end{array}\right. ,
\label{G-U}
\end{eqnarray}
where $U_n(\tau)$ is the proper time evolution operator derived to be
\begin{eqnarray}
 U_n(\vec{x},\vec{y};\tau) 
  &=& \frac1{(4\pi)^{3/2} |\tau|^{3/2}} \frac{QH\tau}{\sin{(QH\tau)}} 
        \exp{\left\{ iQ\int^{\vec{x}}_{\vec{y}} A(\xi)\cdot d\xi \right\}}
\nonumber \\
 && \quad \times \exp{\left[ -\frac{i}4 (x-y)_i QF_{ij} 
                   [\coth(QF\tau)]_{jk} (x-y)_k  \right. }
\nonumber \\
 &&  \left. \hspace{3cm}
    -i\tau \left\{ \frac12 QF_{jk} \sigma_{jk} 
              -(\omega_n -i\mu)^2 - m^2 \right\} \right].
\end{eqnarray}
Substituting Eqs. (\ref{Til-G}) and (\ref{G-U}) into Eq. (\ref{Green-G}),
we obtain the fermion Green function $S(x,y;m)$. It should be noted that
there is a problem of ambiguity in selecting the proper time contour for
$\mu \gtrsim T>0$.\cite{Schm} We must be careful to take the physical
contour. 

Performing the integration over $m$ in Eq. (\ref{v-fer}), we obtain the
explicit form of the effective potential 
\begin{eqnarray}
 V_{\rm eff}(\sigma)
  &=& \frac1{4 G} \sigma^2
  + \frac{e^{-3\pi i/4}}{(4\pi)^{3/2} \beta} \sum_{n=0}^\infty
  \int_{+0}^{\infty}\! d\tau \frac{QH}{ \tau^{3/2}}
      \cot{(QH\tau)}  \nn\\
  && \hspace{4.5cm}
    \times  e^{-i\tau (\omega_n -i\mu)^2}
      \left( e^{-i\tau \sigma^2} -1 \right)+(c.c.)  \ ,
\label{Veff-f}
\end{eqnarray}
where $(c.c.)$ denotes the complex conjugate of the second term. In the
limit $\mu\to0$, the effective potential (\ref{Veff-f}) reproduces the
previous result obtained in Ref. \citen{Ditt}. In the other limit,
$T\rightarrow 0$, Eq. (\ref{Veff-f}) is consistent with the results
given in Refs. \citen{Chod} and \citen{Persson:1994pz}. 

\begin{figure}[t]
 \begin{center}   
    \includegraphics[width=10cm]{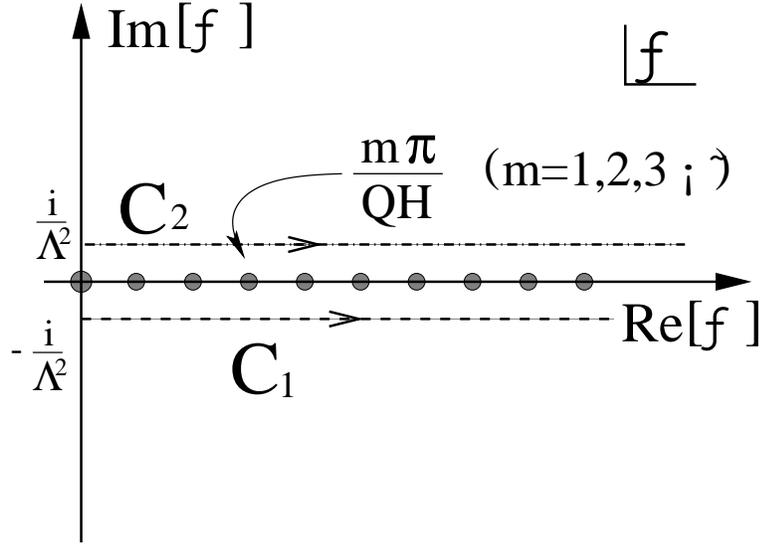}
 \end{center}
\caption{The contours of the integration in Eq. (\ref{Veff-f}). 
 The circles denote the poles that come from $\tau^{-3/2}$ and
 $1/\sin{(QH\tau)}$. $\Lambda$ is a proper time cut-off. }
\label{path}
\end{figure}

The integrand in Eq. (\ref{Veff-f}) has many poles, which come from
$|\tau|^{-3/2}$ and $\cot{(QH\tau)}$ on the real axis. The contour is
determined by the boundary conditions of the Green function. The
physical contour is obtained in Ref. \citen{Schw} for $T=\mu=0$. Because
our result must coincide with that obtained using this contour in the
simultaneous limits $T\to0$ and $\mu\to0$, the contour for the proper
time integral in the second term of Eq. (\ref{Veff-f}) should be $C_1$
in Fig. \ref{path}. We should employ the contour $C_2$ for the complex
conjugate term.  

To determine the contribution from the poles on the real $\tau$ axis, we
rewrite Eq. (\ref{Veff-f}) in the form 
\begin{equation}
 V_{\rm eff}(\sigma) 
  =\frac{\sigma^2}{4G} + I(\sigma) -I(\sigma=0) + I^*(\sigma) -I^*(\sigma=0) ,
\label{Veff-fin}
\end{equation}
where 
\begin{eqnarray}
 I(\sigma) 
   &=& \frac{e^{-3\pi i/4}}{(4\pi)^{3/2} \beta} \sum_{n=0}^\infty
     \int_{C_1}\! d\tau \frac{QH}{\tau^{3/2}}  \cot{(QH\tau)} 
       e^{-i\tau(\omega_n-i\mu)^2 -i\tau\sigma^2} .
\label{I-sig}
\end{eqnarray}
After the Wick rotation in the complex $\tau$-plane, the contribution
from the poles is classified into the following two cases. 
\vspace{0.3cm} \\
{\bf a)} $\omega_0{}^2 + \sigma^2 - \mu^2 
 = (\pi T)^2+\sigma^2-\mu^2>0 $

In this case, the integrand in Eq. (\ref{I-sig}) converges as $\tau\to
-i\infty$. Hence, the expression (\ref{I-sig}) can be continued to
imaginary $\tau$ without encountering any poles:
\begin{eqnarray}
 I(\sigma) +I^*(\sigma)  &=& \frac{1}{4\pi^{3/2}\beta} \sum_{n=0}^\infty 
  \int_{1/\Lambda^2}^\infty d\tau\ f(\sigma;n;\tau)  \ ,
\label{Veff>T} \\
 f(\sigma;n;\tau) &=& \frac{QH}{\tau^{3/2}} \coth(QH \tau)
  \cos(2\omega_n \mu \tau) e^{-\tau (\omega_n{}^2-\mu^2+\sigma^2)} \ .
\end{eqnarray}
We can rewrite the summation in this expression by using the elliptic
theta function, and doing so, we reproduce the results obtained in
Ref. \citen{Kane}. 
\vspace{0.3cm}\\
{\bf b)}  $\omega_0{}^2 + \sigma^2 - \mu^2 
  = (\pi T)^2+\sigma^2-\mu^2<0 $ 

In this case, the integrand in (\ref{I-sig}) is exponentially suppressed
in the limit $\tau\to-i\infty$ for $n>[N]$ and in the limit
$\tau\to+i\infty$ for $n<[N]$, where
$N(\sigma)=(\beta\sqrt{\mu^2-\sigma^2}/\pi-1)/2$ and $[N]$ is the
maximum integer smaller than $N$. Thus, we must add the residues of the
poles for $n<[N]$. This yields
\begin{eqnarray}
 I(\sigma) +I^*(\sigma) &=& \frac{1}{4\pi^{3/2}\beta} \left[
  \sum_{n>[N]}^\infty \int_{1/\Lambda^2}^\infty d\tau  
  f(\sigma;n;\tau) \right. \nn \\
  && \quad
  +\left. \sum_{n=0}^{[N]} \left\{ h_0(\sigma;n)+h_j(\sigma;n)
   +\int_{1/\Lambda^2}^\infty d\tau\ g(\sigma;n;\tau) \right\} \right] , \\
\label{Veff<T}
 g(\sigma;n;\tau) 
 &=& \frac{QH}{\tau^{3/2}} \coth(QH\tau)
  \sin(2\omega_n \mu\tau) e^{\tau (\omega_n{}^2-\mu^2+\sigma^2)} \ , \\
 h_0(\sigma;n) &=& \frac{e^{-i\pi/4}}{2} QH\Lambda
   \int_{-\pi/2}^{\pi/2}d\theta\ e^{-i\theta/2} 
   \cot(QH e^{i\theta} \Lambda^2  ) \nn\\
 &&\hspace{1cm}\times \exp[-i \{ (\omega_n-i\mu)^2+\sigma^2 \}
  e^{i\theta/\Lambda^2} ]+(c.c.) \ , \\
 h_j(\sigma;n) &=& \frac{2}{\sqrt\pi} \sum_{l=1}^{\infty}
  \left( \frac{QH}{l} \right)^{3/2} e^{-2 \pi l\omega_n \mu/(QH)} \nn\\
 &&\hspace{1cm}\times \sin \left\{ \frac{\pi l}{QH} 
  (\omega_n{}^2-\mu^2+\sigma^2) + \frac{3\pi}{4} \right\} \ . 
\end{eqnarray}

By choosing the contour of the proper time integration carefully, we
find explicit expressions for the effective potential with
Eqs. (16)-(21), without any ambiguities, even for $\mu \gtrsim
T>0$. This result is consistent with the previous results in the
simultaneous limits $H \rightarrow 0$ and $1/\Lambda^2 \rightarrow
0$. For example, the effective potential (\ref{Veff-fin}) exactly
reproduces the result given in Ref. \citen{Ina2} at finite $T$ and
$\mu$.\footnote{ To compare the effective potential (\ref{Veff-fin})
with that obtained in Ref. \citen{Ina2}, we must rewrite the equation
using the renormalized coupling constant $G_r$, which is defined by
the renormalization condition $\displaystyle{
\left.\frac{\partial^2 V}{\partial\sigma^2}\right|_{\sigma=\sigma_0}
   = \frac1{G_r}}$.}

\section{Phase structure}
 
In this section, we plot the phase structure of the NJL model at finite
$T$, $\mu$ and $H$. For this purpose, we numerically calculated the
effective potential (\ref{Veff-fin}). The parameters of our model are the 
coupling constant $G$ and the proper time cutoff $\Lambda$. We fix these
parameters to the physical mass scale that reproduces the observed
values. Here, it is assumed that the $T$, $\mu$ and $H$ dependences of
$G$ and $\Lambda$ are not so strong. Below, we use the values $G= 38.7
{\rm GeV^{-2}}$ and $\Lambda=0.864{\rm GeV}$ obtained in the appendix. 
\begin{figure}[ht]
\begin{center}
 \begin{tabular}{cc}
    \resizebox{!}{4.6cm}{\includegraphics{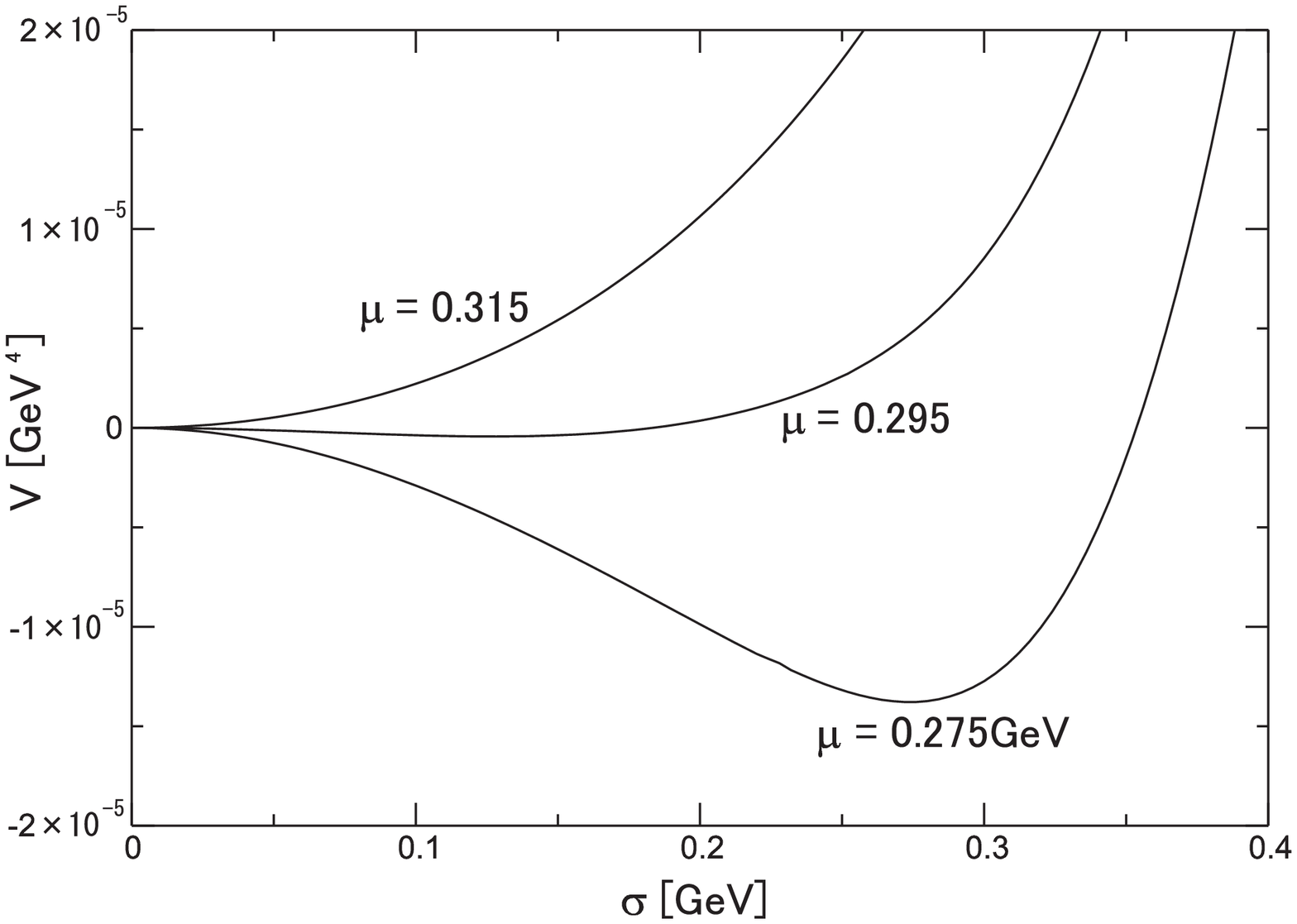}}
    & 
    \resizebox{!}{4.6cm}{\includegraphics{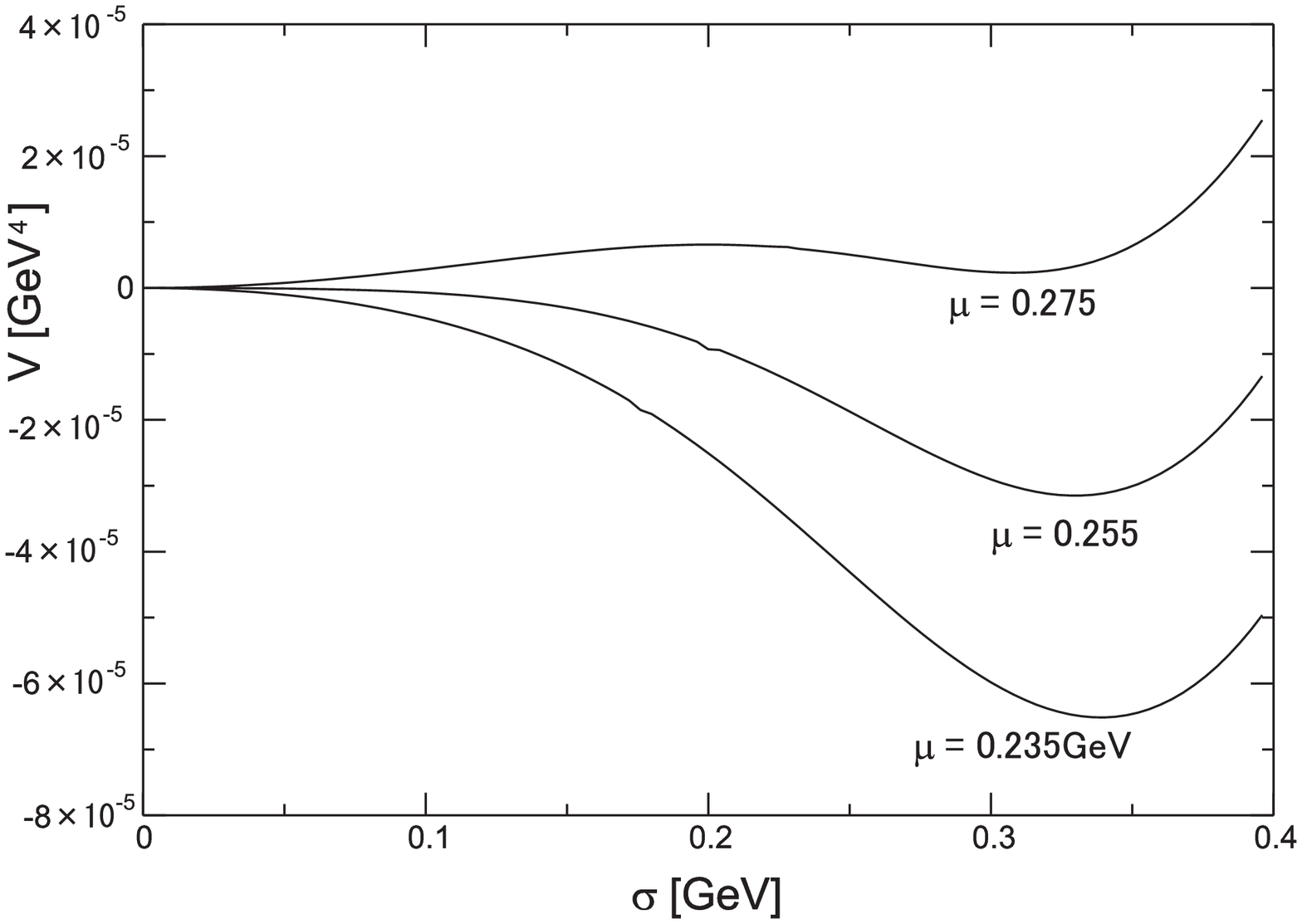}}
    \\
     {(a) At $T=0.05~{\rm GeV}$, $H=0.18~{\rm GeV^2}$. }
    & 
     {(b) At $T=0.05~{\rm GeV}$, $H=0.3~{\rm GeV^2}$. } 
 \end{tabular}
    \resizebox{!}{5.2cm}{\includegraphics{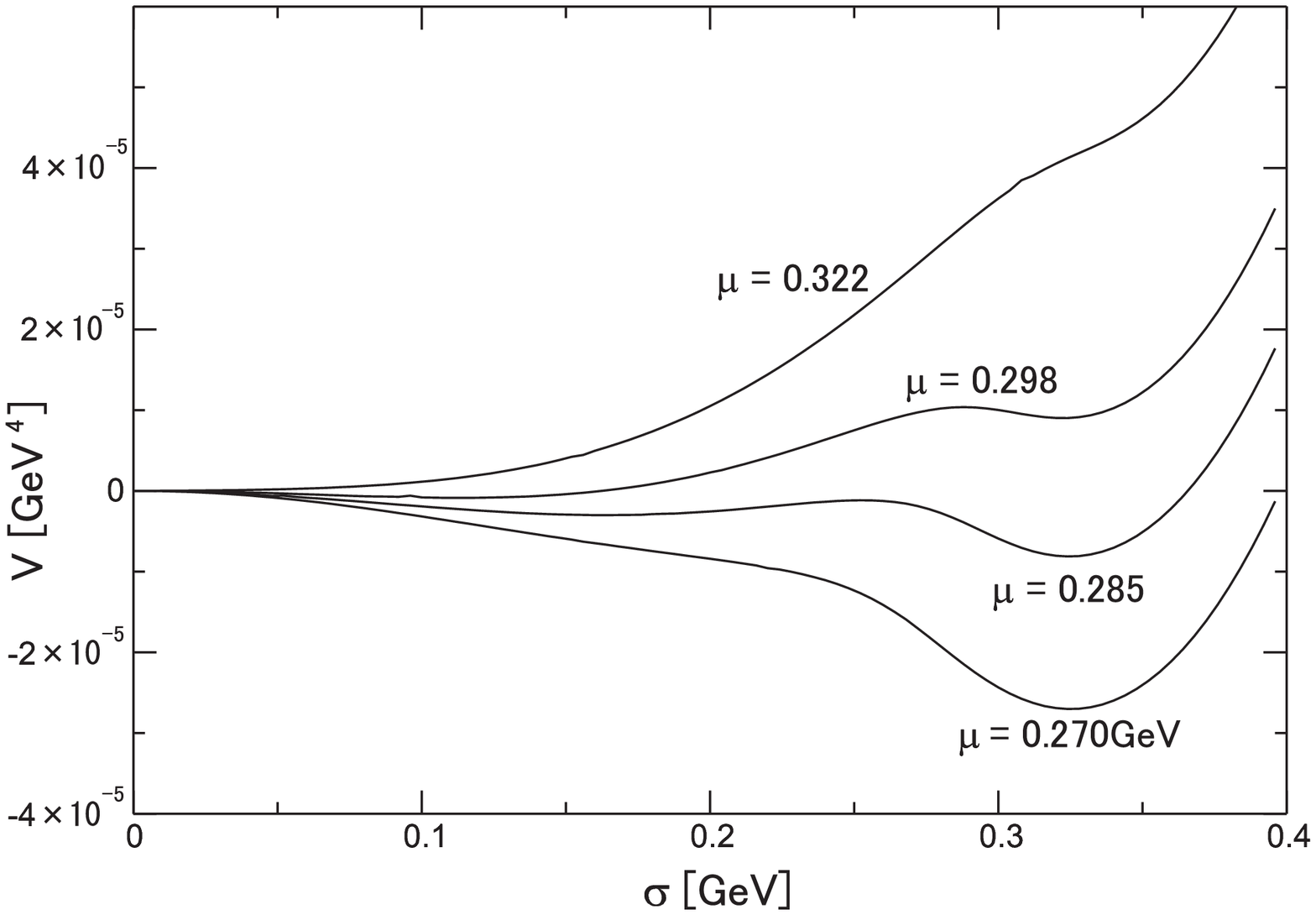}} 
    \\
     {(c) At $T=0.01~{\rm GeV}$, $H=0.18~{\rm GeV^2}$.} 
 \caption{Behavior of the effective potential with $T$, $\mu$ and $H$
   fixed.} 
 \label{pot}
\end{center}
\end{figure}

Typical behavior of the effective potential is plotted in
Fig.~\ref{pot}. As seen in the figures, we observe the followings:\\
(a) There is a second order phase transition as $\mu$ is increased,
  with $T$ kept large and $H$ kept small. \\
(b) There is a first order phase transition  
  as $\mu$ is increased, with $H$ kept large and fixed. \\
(c) There are two steps of the transition as $\mu$ is increased,
  with $T$ and $H$ kept small.\footnote{Such a two-step
  transition is also found in the non-Abelian gauge theory in the
  Randall-Sundrum background.\cite{Abe}}  
  First, a first order transition takes place from the third non-vanishing
  extremum to the first extremum. Next, a second order phase transition
  occurs where the first non-vanishing extremum disappears.

\begin{figure}[t]
  \begin{center}
    \includegraphics[width=13cm]{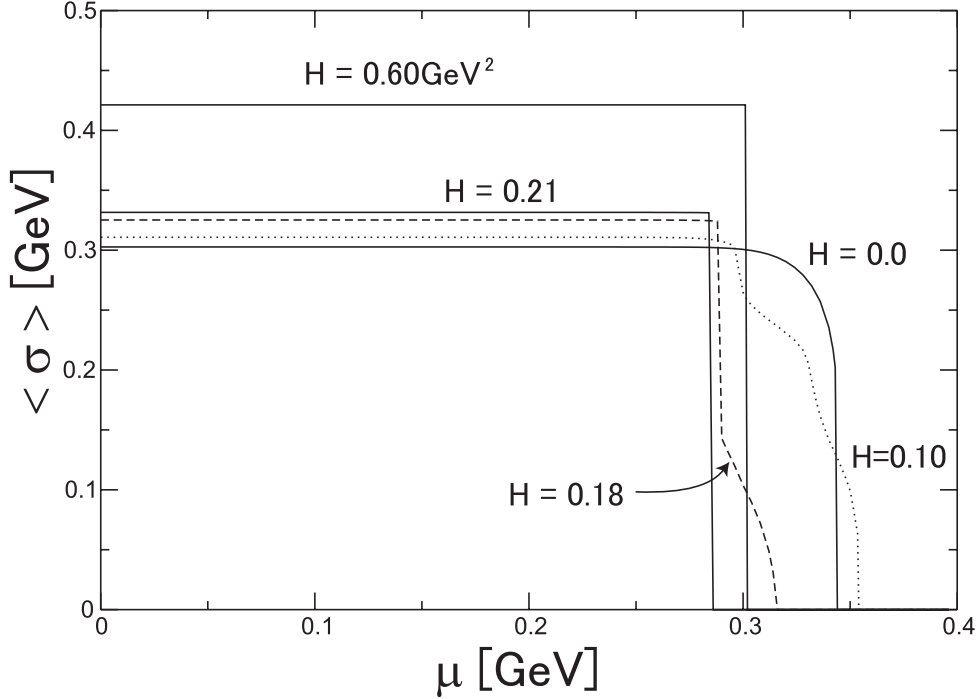}
  \end{center}
\caption{Dynamical fermion mass $\langle \sigma \rangle$ as a function
 of the chemical potential $\mu$ at $T=0.01~{\rm GeV}$ with $H$ fixed at
 the typical values $0.0$, $0.10$, $0.18$, $0.21$ and $0.60~{\rm GeV^2}$.}
\label{fig-d}
\end{figure}
The ground state is determined by finding the minimum of the effective
potential. The necessary condition for this minimum is given by the gap
equation 
\begin{eqnarray}
 \left.\frac{\partial V}{\partial \sigma} 
    \right|_{\sigma=\langle\sigma\rangle} =0 ,
\label{Gap}
\end{eqnarray}  
where $\langle\sigma\rangle$ is the expectation value of $\sigma$, which
corresponds to the dynamically generated fermion mass. If the gap
equation (\ref{Gap}) has more than one solution, we numerically
calculate the effective potential for each of these solutions and
thereby find the value $\langle\sigma\rangle$ in the ground state. In
Fig.~\ref{fig-d}, we plot the behavior of the dynamical fermion mass
$\langle\sigma\rangle$. We find that $\langle\sigma\rangle$ disappears
above a certain critical chemical potential. We clearly observe that
$\langle\sigma\rangle$ increases continuously as $H$ increases for a
small chemical potential. The magnetic field enhances the chiral
symmetry breaking. This phenomenon is known as
$magnetic~catalysis$\cite{Suga,Mira:1994PRL}. For large $\mu$, the
behavior of the dynamical fermion mass is different from that in the
case of magnetic catalysis. Near the critical chemical potential, a
variety of phenomena are observed. For $H=0.10~{\rm GeV^2}$, there is an
oscillation of the critical line. This is called the de Haas-van Alphen
effect\cite{Eber1,Shar}. It is caused when the Landau levels pass the
quark Fermi surface. If $H$ is large enough, the critical chemical
potential increases as $H$ increases, and a mass gap exists at the
critical value of $\mu$. However, there is a complex behavior of the
critical chemical potential for sufficiently small $H$. We observe the
two steps of the transition at $H=0.18~{\rm GeV}^2$. 

\begin{figure}[ht]
  \begin{center} 
   \begin{tabular}{cc}
    \resizebox{!}{4.6cm}{\includegraphics{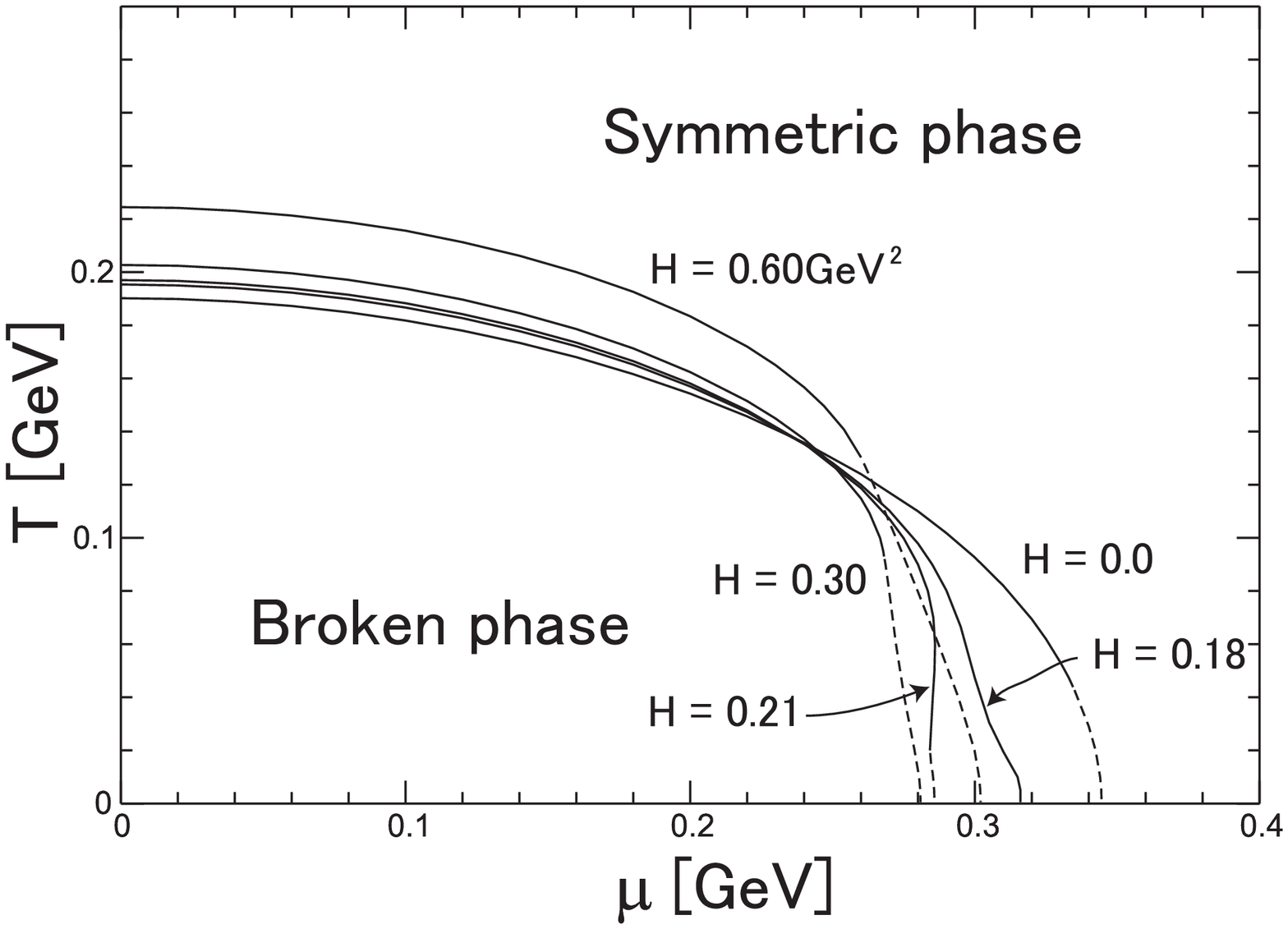}} 
    & 
    \resizebox{!}{4.6cm}{\includegraphics{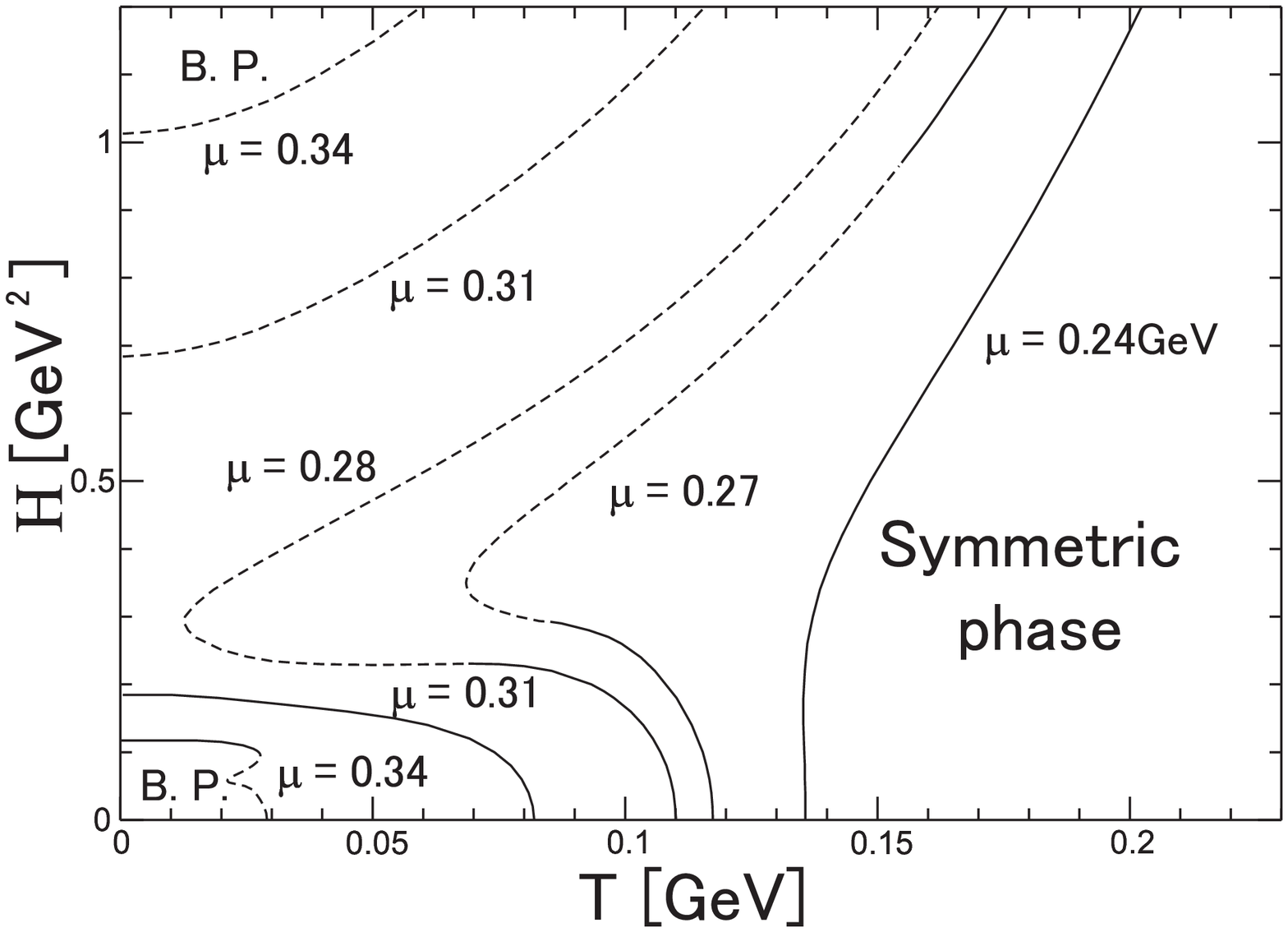}} \\
    {(a) The $T$-$\mu$ plot with $H$ fixed.}   
    & 
    {(b) The $H$-$T$ plot with $\mu$ fixed.}
   \end{tabular}
   \resizebox{!}{4.6cm}{\includegraphics{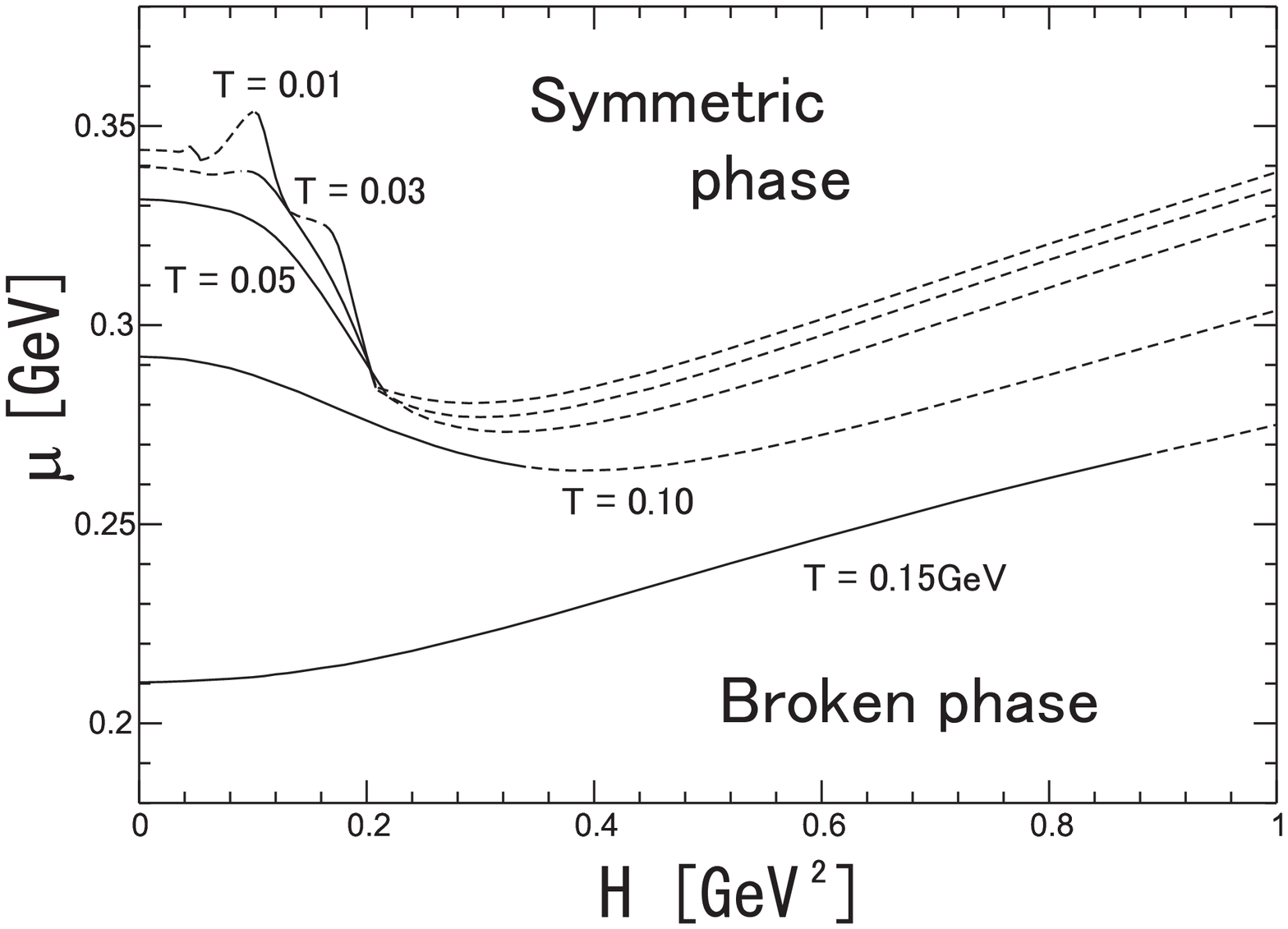}} \\
   {(c) The $\mu$-$H$ plot with $T$ fixed.}
  \end{center}
\caption{Critical lines in the $T$-$\mu$ plane, $H$-$T$ plane and
 $\mu$-$H$ plane. The dashed curve represents the first order phase
 transition, while the solid curve represents the second order phase
 transition.} 
  \label{plane}
\end{figure}

To understand the situations more precisely, we numerically calculated
the critical values of $T$, $\mu$ and $H$ at which the dynamically
generated fermion mass disappears. For the second order phase transition,
the critical point is obtained by taking the following limit of the gap
equation:
\begin{eqnarray*}
 \lim_{\langle \sigma \rangle\to 0} \left\{~ \left. 
    \frac{\partial V}{\partial\sigma}
   \right|_{\sigma=\langle \sigma \rangle }\right\} =0 .
\end{eqnarray*}
For the first order phase transition, we directly observe the behavior of
the effective potential and find the critical point. We plot the phase
boundaries in Fig.~\ref{plane}. On the $T$-$\mu$ plane, it is clearly
observed that the broken phase spreads out as $H$ is increased for
a small chemical potential. The first order phase transition disappears
near $H=0.18~{\rm GeV^2}$.  

We illustrate the details of the phase boundaries at large chemical
potential in the $H$-$T$ and $\mu$-$H$ planes. It is seen that the first
and second order phase transitions coexist in the planes. As is seen in
Fig.~\ref{plane}(b), the broken phase is separated into two parts for
$\mu> 0.28~{\rm GeV}$. The distortion of the critical line at $\mu=
0.34~{\rm GeV}$ is caused by the de Haas-van Alphen effect. Such an
effect is observed as an oscillating mode for $T\lesssim 0.03~{\rm GeV}$
in the $\mu$-$H$ plane in Fig.~\ref{plane}(c). In the parameter range
corresponding to the second order phase transition, chiral symmetry
breaking is suppressed as $H$ increases, while it is enhanced in the
range corresponding to the first order phase transition. 

\section{Summary}

We investigated the phase structure of the NJL model at finite
temperature and chemical potential in an external magnetic field. The
Fock-Schwinger proper time method was applied to a thermal field
theory. Using the two-point function obtained in Ref.~\citen{IKM}, we
calculated an effective potential that exactly includes the effects of
temperature, chemical potential and constant magnetic field to leading
order in the $1/N_c$ expansion. We determined the physical contour of
the proper time integral. There is no ambiguity with regard to the
contour of the proper time integration for $\mu \gtrsim T>0$ in the
expression of the effective potential. 

We used the current algebra relationship with the current quark mass to
determine the mass scale of the model. The position of the tricritical
point depends strongly on the mass scale and the regularization
procedure. Here, we used the parameter values $G=38.7~{\rm GeV^{-2}}$
and $\Lambda=0.864~{\rm GeV}$, where the tricritical point appears in
the $T$-$\mu$ plane.   

Through the study of the shape of the effective potential, we observed 
a phase transition from the broken phase to the symmetric phase when 
the temperature $T$, chemical potential $\mu$ and magnetic field $H$ 
vary. Increasing the temperature and chemical potential tends to prevent 
the chiral symmetry breaking at $H=0$, while increasing the magnetic
field enhances the symmetry breaking at $T=\mu=0$.  Therefore, complex
behavior of the phase boundary is found with the combined effects of
$T$, $\mu$ and $H$, as shown in Fig.~3. We found a two-step transition
from the broken phase to the symmetric phase for a medium size magnetic
field, $H\sim 0.18~$GeV${}^2$. We found a separation of the broken 
phase into two parts in the $H$-$T$ plane. With this separation, two
tricritical points appear in the critical line for $\mu \sim
~0.27$~GeV. More than two tricritical points were observed in the
critical line for $T \sim 0.01$~GeV in the $\mu$-$H$ plane. In the phase
diagram, we observed a contribution from the magnetic field, which is
known as magnetic catalysis, and the de Haas-van Alphen effect.

In the present paper, we regard a simple four-fermion interaction 
model as a low energy effective theory that may possess some fundamental
properties of the chiral symmetry breaking in QCD. Although the present
work is restricted to an analysis of the phase structure of the
four-fermion interaction model with massless quarks, we hope to apply
our result to critical phenomena in the real world. In QCD, the current
quark mass for up and down quarks is much smaller than the constituent
quark mass $\sigma$ for $T=\mu=H=0$. Because the expectation value of
$\sigma$ decreases near the critical point, it is no longer valid to
ignore the current quark mass. If the contribution from the current
quark mass is included in the effective potential, the sharp transitions
illustrated in Fig.~4 disappear, as shown in Refs.\citen{Suga} and
\citen{Hats:74}. In that case, the second order phase transition becomes
a cross over.  

The strength of the magnetic field that can affect the phase structure
is larger than that existing in the core of neuron stars and
magnetars. However, an oscillating mode has been observed even for small
magnetic fields, and this may have some interesting contribution to the
behavior of such stars. Furthermore, we believe that the magnetic field
may have been significantly stronger in the early stages of our universe
with the primordial magnetic field and that this may have contributed to
the critical phenomena in the early universe. To investigate the
phenomena exhibited by dense stars, we cannot avoid considering the
vortex configurations that generate non-local ground state in the
super-fluidity phase. It is necessary to calculate the effective action
at finite $\mu$ and $H$. We will continue our work further and extend
our analysis to non-local ground states. 

\section*{Acknowledgements}
The authors would like to thank Takahiro Fujihara and Xinhe Meng for
helpful discussions. We are also grateful for useful discussions at the
Workshop on Thermal Quantum Field Theory and Their Applications, held 
at the Yukawa Institute for Theoretical Physics, Kyoto, Japan, August 
2003.

\appendix
\section{Physical Scale of $G$ and $\Lambda$}

Here we discuss the physical scale of the parameters $G$ and $\Lambda$
of our model. We regard the NJL model as a low energy effective theory
of QCD. The parameters should be determined so as to reproduce the
observed quantities at low energy scale.

First, we calculate the pion decay constant using the proper time
method. The two-point function for a pion composite field is given by 
\begin{eqnarray}
 \Gamma_\pi^{(2)}(p) &=& -\frac{N_c}{G} +iN_c \int\frac{d^4 k}{(2\pi)^4}
  \int d^3\vec{x} d^3\vec{x}'\ e^{-i\vec{k}\cdot\vec{x}}
  e^{-i(\vec{k}+\vec{p})\cdot\vec{x}'} \nn\\
  &&\hspace{2cm} \times {\rm tr}
  [(-i\gamma^5 \tau^a)S(\vec{x},0;k_0)
             (-i\gamma^5 \tau^b)S(\vec{x}',0;k_0+p_0)] ,
\label{Gamm-pi}
\end{eqnarray}
where $S(\vec{x},\vec{y};k_0)$ is the quark Green function defined by 
\begin{eqnarray}
  (\gamma^0 k_0+i\gamma^j\partial_j-\sigma) S(\vec{x},\vec{y};k_0)
  = \delta^3(\vec{x}-\vec{y})  .
\end{eqnarray}
Here, $\sigma$ is determined by finding the stationary point of the
effective potential which is determined as a function of $G$ and
$\Lambda$. Using the proper time method, the Green function
$S(\vec{x},\vec{y};k_0)$ can be written 
\begin{equation}
 S(\vec{x},\vec{y};k_0) 
    = \frac{e^{\pi i/4}}{(4\pi)^{3/2}} 
       ( i\gamma^j\partial_j+ \gamma^0 k_0 +\sigma)
     \int^{-0}_{-\infty}\! \frac{d\tau}{\tau^{3/2}} 
      \exp{\left\{ -\frac{i}4 \frac{(\vec{x}-\vec{y})^2}\tau 
           +i(\sigma^2 -k_0^2)\tau \right\}}   ,
\label{Qua-G}
\end{equation}
where $\tau$ is the proper time. Substituting Eq.~(\ref{Qua-G}) into
Eq.~(\ref{Gamm-pi}), and performing the integration over $k$, $x$ and
$x^\prime$, we obtain  
\begin{eqnarray}
  \Gamma_\pi^{(2)}(p)
   &=&  -\frac{N_c}{G} + \frac{N_c}{2\pi^2} \int^\infty_{+0} d\tau ds\ 
   e^{-i\sigma^2(\tau+s)}e^{ip^2\tau s/(\tau+s)} \nn\\
  &&\hspace{2cm} \times\frac{1}{(\tau+s)^2} \left[ \frac{1}{\tau+s}
   \left\{ 2i - p^2 \frac{\tau s}{\tau+s} \right\} - \sigma^2 \right] .
\label{Gamma} 
\end{eqnarray}
The renormalization constant for the pion wave function is defined 
by $\Gamma_\pi^{(2)}(p^2) = Z_\pi^{-1} p^2+O(p^4)$. After applying the
Wick rotation to Eq.~(\ref{Gamma}) the renormalization constant $Z_\pi$
is derived as
\begin{eqnarray}
 Z_\pi^{-1}=\frac{N_c}{2\pi^2}\int^\infty_{+0} d\tau ds\ 
  e^{-\sigma^2(\tau+s)}\frac{\tau s}{(\tau+s)^3} \left(
  \frac{3}{\tau+s}+\sigma^2 \right) \ .
\label{Z-pi}
\end{eqnarray}

To obtain the relationship between $f_\pi$ and $Z_\pi$, we evaluate the
order parameter of the chiral symmetry breaking, $[iQ_5^a,\pi^b]$, where
$Q_5^a$ is the conserved axial charge and $\pi^a$ is the asymptotic pion
field. Then, using the conserved axial current $j_\mu^a(x)$, we obtain
\begin{eqnarray}
 \left< 0\left| [iQ_5^a,\pi^b] \right|0\right>
  &=& \langle 0 | [ \int\!d^3\vec{x}~j^a_0(x) ,\pi^b ] | 0 \rangle
\nonumber \\
 &=& \int\! d^4x ~i\partial^\mu 
      \left< 0\left|  T j^a_\mu(x) \pi^b \right|0\right> .
\label{Curr1}
\end{eqnarray}
The pion decay constant $f_\pi$ is defined by
\begin{equation}
\left< 0 \left| j^a_\mu (x) \right| \pi(p) \right> 
= -if_\pi\delta^{ab} p_\mu e^{-ipx} .
\label{dconst}
\end{equation}
This implies that 
\begin{equation}
 j^a_\mu (x)  = f_\pi \partial_\mu \pi^a(x) + \cdots .
\label{Curr2}
\end{equation}
Substituting Eq.~(\ref{Curr2}) into Eq.~(\ref{Curr1}), we obtain 
\begin{equation}
 \left< 0\left| [iQ^a_5,\pi^b] \right|0\right>  = f_\pi \delta^{ab}.
\label{Curr3}
\end{equation}
On the other hand, $j^a_\mu(x)$ and $\pi^a(x)$ are described by the quark
field $\psi(x)$ as
\begin{equation}
 j^a_\mu(x) = \bar\psi \gamma_\mu \gamma_5 \frac{\tau^a}2 \psi  
      \quad ,\quad 
    \pi^a(x)=  -\frac{G}{N_c} Z_\pi^{-1/2} \bar\psi i \gamma_5 \tau^a \psi .
\label{Curr4}
\end{equation}
Inserting Eq.~(\ref{Curr4}) into the second line of Eq.~(\ref{Curr1}),
we find
\begin{equation}
 \left< 0\left| [iQ^a_5,\pi^b ] \right|0\right>  
   =  -\frac{G}{N_c} Z_\pi^{-1/2} 
         \left< 0\left| -\bar\psi \psi \right|0\right> \delta^{ab}
   = - Z_\pi^{-1/2} \sigma \delta^{ab} .
\label{Curr5}
\end{equation}
Comparing (\ref{Curr3}) with (\ref{Curr5}), we obtain the relationship
$f_\pi=-Z_\pi^{-1/2} \sigma $. From Eq.~(\ref{Z-pi}) and this
relationship, it is found that the pion decay constant can be expressed
as a function of $\sigma$ and $\Lambda$,  
\begin{eqnarray}
 f_\pi{}^2 = \frac{N_c \sigma^2}{2\pi^2} \int_{1/\Lambda^2}^\infty d\tau 
   \frac1{2\tau}  \left\{
      e^{-\sigma^2 \tau}  (1-\sigma^2\tau) 
  - \sigma^4 \tau^2 {\rm Ei}(-\sigma^2\tau) \right\} ,
 \label{f_pi}
\end{eqnarray}
where ${\rm Ei}(-x)$ is the exponential-integral function.

\begin{figure}[t]
 \begin{center}   
    \includegraphics[width=10cm]{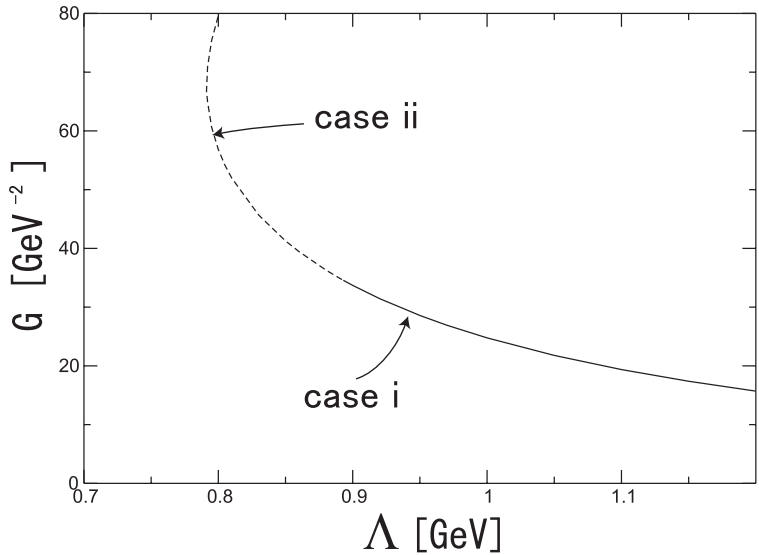}
 \end{center}
\caption{Relationship between $G$ and $\Lambda$.} 
\label{G-Lam}
\end{figure}

\begin{figure}[ht]
  \begin{center}
    \includegraphics[width=13cm]{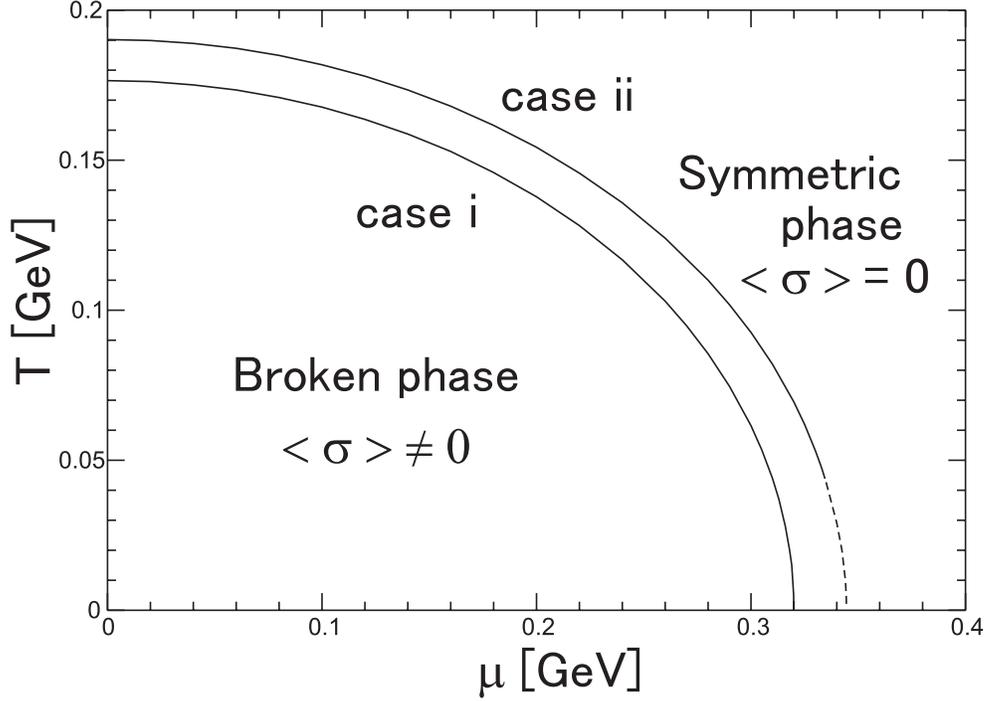}
  \end{center}
\caption{Critical lines at $H=0$ in the cases {\bf i} and {\bf ii}.
 The dashed curve represents the first order phase transition, while the
 solid curve represents the second order phase transition.} 
\label{mass-diff}
\end{figure}

We determine the relationship between $G$ and $\Lambda$ such that the
measured value, $f_\pi=93~{\rm MeV}$, is reproduced, as shown in 
Fig.~\ref{G-Lam}.
The phase structure in the $T$-$\mu$ plane differs dramatically 
for values of $G$ and $\Lambda$ on the solid curve (case {\bf i}) and
the dashed curve (case {\bf ii}) in Fig.~\ref{G-Lam}. Evaluating the
effective potential, we plot the typical behavior of the critical line
in $T$-$\mu$ plane in both cases. As shown in Fig.~\ref{mass-diff}, only
the second order phase transition takes place in case {\bf i}, while the
first and second order phase transitions coexist in case {\bf ii}, and a
tricritical point appears on the critical line. These results agree with
the well-known result obtained using the other regularization
procedure.\cite{Hatsu,Ina2,Fod} Therefore, we choose the mass scale in a
range corresponding to case ${\bf ii}$. In chiral limit, we determine
only the curve in the $G$-$\Lambda$ plane.  

To determine the values of $G$ and $\Lambda$ in this range, we evaluate
the pion mass. As is well known, the pion is a massless Goldstone mode
if the explicit chiral symmetry breaking is not accounted for. In QCD, a 
small quark mass breaks the chiral symmetry. The pion decay constant 
is proportional to $\sigma$. Because the value of $\sigma$ is much larger 
than the current quark mass, we ignore the a contribution from the
current quark mass when calculating the pion decay constant. However, to
evaluate the realistic pion mass, $m_\pi = 138~{\rm MeV}$, we consider
the influence of the current quark mass. 

Accounting for the current quark mass, the axial current is not
conserved, and we have 
\begin{equation}
\partial^\mu j^a_\mu (x) = \hat{m} \bar{\psi}i\gamma_5\tau^a\psi(x),
\label{acurrent}
\end{equation}
where $\hat{m}$ is the current quark mass, $2\hat{m} = m_u+m_d$.
Substituting Eq.~(\ref{dconst}) into Eq.~(\ref{acurrent}), we obtain
\begin{equation}
\widehat{m} 
\left< 0 \left| \bar{\psi}i\gamma_5\tau^a\psi(0)\right| \pi^b(p=0) \right>
= -f_\pi m_\pi^2 \delta^{ab}.
\label{ac2}
\end{equation}
Similarly to Eqs. (\ref{Curr1}) and (\ref{Curr3}), we obtain
\begin{equation}
-f_\pi \left< 0 \left| \bar{\psi}i\gamma_5\tau^a\psi(0)\right| 
       \pi^b (p=0) \right>
= \left< 0 \left| \bar{\psi}\psi \right| 0 \right> 
\end{equation}
in the chiral limit. Therefore, the pion mass is obtained as
\begin{equation}
 (f_\pi m_\pi)^2 = -\frac{N_c}G ~\hat{m}\langle\sigma\rangle ,
\label{fm}
\end{equation}
to first order in the current quark mass.\cite{Hats:74} This is known as
the Gell-Mann$-$Oakes$-$Renner relation. 

From the measured values $f_\pi = 93~{\rm MeV}$ and $m_\pi = 138~{\rm
MeV}$ and Eqs.~(\ref{f_pi}) and (\ref{fm}), we obtain the matching
condition for the coupling constant $G$ and the proper time cutoff
$\Lambda$ with the physical scale. These parameters are extracted as
functions of $\hat{m}$. For $\hat{m}=5.5~{\rm MeV}$, the parameters $G$
and $\Lambda$ are found to be 
\begin{equation}
 G= 25.4~{\rm GeV^{-2}} \quad ,\quad \Lambda = 0.991~{\rm GeV} ,
\end{equation}
which are located in the range of the solid curve in Fig.~\ref{G-Lam}.
A larger current quark mass is realized on the dashed curve in
Fig.~\ref{G-Lam}. If we set $\hat{m}=7.0~{\rm MeV}$ in Eq.~(\ref{fm}),
the parameters $G$ and $\Lambda$ are fixed as 
\begin{equation}
G= 38.7~{\rm GeV^{-2}} \quad ,\quad \Lambda = 0.864~{\rm GeV} ,
\end{equation}
which are in the range of the dashed curve in Fig.~\ref{G-Lam}. 

It should be noted that we consider the contribution from the current
quark mass only for the purpose of determining the values of $G$ and
$\Lambda$.

\end{document}